\documentclass[twocolumn,showpacs,preprintnumbers,amsmath,amssymb]{revtex4}

\newcommand{\be}{\begin{equation}}
\newcommand{\ee}{\end{equation}}
\newcommand{\beq}{\begin{eqnarray}}
\newcommand{\eeq}{\end{eqnarray}}

\usepackage{graphicx}
\usepackage{dcolumn}
\usepackage{bm}

\begin{document}
\def\ua{\uparrow}
\def\da{\downarrow}\def\ra{\rightarrow}
\def\a{\alpha}
\def\b{\beta}
\def\g{\gamma}
\def\G{\Gamma}
\def\d{\delta}
\def\D{\Delta}
\def\e{\epsilon}
\def\ve{\varepsilon}
\def\z{\zeta}
\def\h{\eta}
\def\th{\theta}
\def\l{\lambda}
\def\L{\Lambda}
\def\m{\mu}
\def\n{\nu}
\def\x{\xi}
\def\X{\Xi}
\def\p{\pi}
\def\P{\Pi}
\def\r{\rho}
\def\s{\sigma}
\def\S{\Sigma}
\def\t{\tau}
\def\f{\phi}

\preprint{APS/123-QED}

\title{Microscopic Model for a Strongly Correlated Superconducting
Single-Electron-Transistor}

\author{Enrico Perfetto  }
\author{Michele Cini}%
 \email{Michele.Cini@roma2.infn.it}
\affiliation{%
Dipartimento di Fisica, Universita' di Roma Tor Vergata, Via della
Ricerca Scientifica, 1-00133 Roma, Italy and \\ INFN - Laboratori
Nazionali di Frascati, Via E.Fermi 40, 00044 Frascati, Italy
and\\ Istituto Nazionale per la Fisica della Materia\\
}%


\date{\today}

\begin{abstract}
We model    a  Superconducting Single-Electron
 Transistor  operating by  repulsive
 interactions.
 The device consists of a ring of Hubbard clusters, placed between electrodes
 and capacitively coupled to a gate potential. In each cluster, a
 pair of electrons at appropriate filling feels a weak effective
 interaction which leads to pairing in part of the parameter
 space.
  Thus, the system can host many bound pairs, with correlation induced binding.  When the charging energy exceeds the
  pairing energy, single-electron tunneling prevails;
   in the opposite regime, we predict  the Coulomb blockade pattern of two-electron tunneling.
   This suggests that in tunneling experiments
     repulsion-induced pairs may  behave in a similar way as phonon-induced ones.
\end{abstract}

\maketitle

\section{Introduction}
\label{sec0}

  In recent years, a variety of transport experiments have
been reported in molecular size systems, such as quantum dots and
nanotubes,  as a contribution to the current boost towards of the
progress in nanoscale technology. From the theoretical side,
circuits of several kinds have been modeled\cite{aligia}, and in
applied electronics the
Single-Electron-Transistors\cite{revmodphys} are among the most
important devices. These are realized by connecting a nanoscopic
conducting island to metallic leads and to a gate voltage. The
energy gaps existing between states with different number of
particles allow
 to fix the number of electrons in the island very sharply; as a
 consequence single electrons can tunnel to or from the conductor. Even more
appealing situations  arise when the above scenario is complicated
by electron-electron interactions, as in the case of a
Superconducting-Single Electron Transistor (S-SET).

A S-SET is a mesoscopic device obtained by linking capacitively  a
superconducting grain to two normal leads and to a gate electrode
as well\cite{definizione}. The latter allows one to control the
number $N$ of electrons on the  grain by tuning the gate voltage
$V_{g}$. Such a system has been studied both
experimentally\cite{devoret} and
theoretically\cite{averin}\cite{hekking}\cite{hanke} in great
detail during the past years. In a normal island the parity of $N$
oscillates between even and odd values, by varying $V_{g}$;
conversely in a superconducting island $N$ is always even because
of the paired nature of the ground state.  Therefore the S-SET
transport properties in the linear regime are governed by Andreev
reflection under the critical temperature $T_{C}$ of the central
island, while  above $T_{C}$ single electron tunneling prevails.
This leads to well pronounced Coulomb blockade peaks of the
conductance $G=\partial I / \partial V|_{V=0}$ as a function of
the gate voltage. In particular the parity-controlled tunneling
produces $2e/C_{g}$ periodic peaks in the pair-tunneling regime,
in contrast with the $e/C_{g}$ periodicity of the normal system
(here $C_{g}$ is the capacity of the gate electrode). This
behavior is well reproduced by models
\cite{averin}\cite{hekking}\cite{hanke} using a gate controlled
BCS Hamiltonian $H_{BCS}$; the connection to  free electron leads
employs a tunneling Hamiltonian, usually treated by second-order
perturbation theory.

In the present article we propose a model for a S-SET with a strongly
correlated, repulsive Hubbard-like model instead of $H_{BCS}$ as
the ``superconducting'' grain Hamiltonian. That is, we look for a
superconducting response entirely driven by the electronic
correlations rather than by the phonon-mediated effective
attraction. The occurrence of two-electron tunneling in non BCS
systems was observed by  Ashoori {\it et al.}\cite{ashoori} in a
$1\mu$m GaAs tunnel capacitor. Purely electronic mechanisms were
proposed to explain this behavior and the GaAs quantum dot models
ranged from a semiclassical description\cite{Raikh} to a Hubbard
model framework\cite{canali}. Unlike the systems considered by
Refs.\cite{Raikh} \cite{canali}, in our {\em gedankenexperiment },
like in a S-SET, the tunneling current is due to many bound pairs
hosted by the device  in a wide range of gate potentials.

The plane of the paper is the following. In the next Section we
introduce the microscopic model that we are going to study.
Section \ref{sec2} is devoted to determine some important
properties of the  strongly correlated central island. We show
that the electronic correlations provide a non-trivial
characteristic energy which can be compared with the electrostatic
charging energy in order to distinguish between a {\it normal}
regime and a {\it superconducting} one. In particular in these two
regimes  the parity of the number of particles in the ground state
oscillates exactly like in a S-SET. In Section \ref{sec3} we
explicitly calculate the conductance as a function of the gate
voltage by using a master equation approach\cite{beenakker}. It is
found that  the linear  response of our strongly correlated device
shows Coulomb blockade pattern. A normal behavior is observed in
the non-correlated and in the very strongly correlated regimes;
while in the intermediate case, the   spacing between  the
conductance peaks doubles. Finally the conclusions are drawn in
Section \ref{sec4}.

\section{The Model}
\label{sec1}

Let us consider the grand-canonical Hamiltonian
\begin{equation}
    {\cal H}-\mu \hat{N}_{tot}=H_{\rm device}+H_{\rm leads}+H_{T} \, .
    \label{model}
\end{equation}
Here $H_{\rm device}$ is an extended Hubbard model of the central
island coupled capacitively to a gate voltage $V_{g}$; $H_{\rm
leads}$ describes the left and right reservoirs, supposed to be
identical free electron gases for simplicity; $H_{T}$ is the
tunneling Hamiltonian, that connects the central device to the
leads; $\mu$ is the chemical potential and $\hat{N}_{tot}$ is the
total number of particles operator. The carriers are electrons, of
charge $-e, e>0$. Let us examine these three terms  in detail.
\begin{figure}
    \includegraphics[width=8cm]{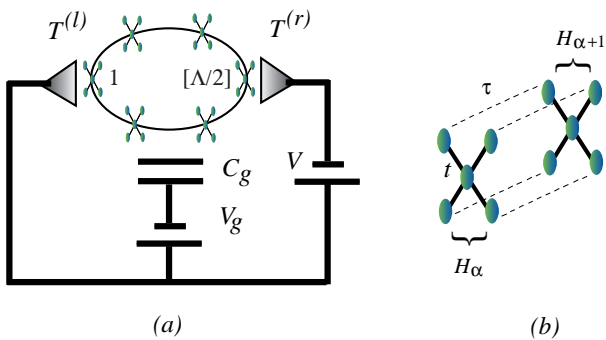}
    \caption{\label{device}$(a)$: Scheme of the strongly correlated S-SET. The device consists
    of $\L$ Hubbard clusters arranged in a ring and linked
    symmetrically to one another.
    $(b)$: Pictorial representation of $H_{\alpha}$
    and $H_{\tau}$. }
\end{figure}
As in previous work\cite{PRB2003}, the central island consists in
a ring of $\L$  identical 5-site centered-square Hubbard clusters,
denoted by the index $\a$, see Figs.\ref{device}$a,b$. Each
cluster is described by the  Hamiltonian
\begin{equation}
 H_{\a} =\sum_{\s,i=1}^{4}t( p^{\dag}_{\a,0 \s}p_{\a, i\s}+
{\rm h.c. })+ U\sum_{i=0}^{4}\hat{n}_{\a, i \ua}\hat{n}_{\a, i
\da} \; , \label{unpham}
\end{equation}
with the creation operators  on the   $\a$-th cluster
$p^{\dag}_{\a,0 \s}$ for the central site, and   $p^{\dag}_{\a,
i\s}$, $i=1,..,4$ for the remaining 4 sites; $\hat{n}_{\a, i
\s}=p_{\a, i \s}^{\dagger}p_{\a, i \s}$, and $\s$ is a spin index.

In the device, each cluster $\a$ is linked to the two nearest
neighbors ones (denoted by $\a +1$ and $\a -1$) by the hopping
Hamiltonian $H_{\t}$ (see Fig.\ref{device}$a$) whereby   a
particle in the $i$-th site of the $\a$-th cluster can hop towards
the $i$-th site of the $\b = \a \pm 1 $-th clusters:
\begin{equation}
H_{\tau}= \t \sum_{ \a  =1}^{\L} \sum_{ \b  =\a \pm 1}
\sum_{\s,i=1}^{4}   (  p_{\a, i\s}^{\dag} p_{\b, i\s}  + {\rm
h.c.})\; . \label{htau}
\end{equation}
$H_{\rm device}$ also  contains  an electrostatic charging energy
term due to an effective capacitance $C$ of the central island.
Finally the island is connected capacitively to the gate which is
at a potential $V_{g}$ (see Fig.\ref{device}$b$). Therefore we
have
\begin{equation}\label{accac}
 H_{\rm device}= \sum_{\a =1}^{\L}H_{\a}+ H_{\t}+ \frac{(\hat{N}e)^{2}}{2C} -e(V_g-\mu) \hat{N}
\end{equation}
where $\hat{N}$ is total number of particles operator in the
central device. We remark that  the  capacitive term is
essentially long-ranged and accounts for the monopole contribution
to the charging energy, while the $U$ terms depend on the way the
charges are distributed in the island.
 In   all
electrostatic terms $\hat{N}$ should be  referred to an average
population corresponding to a neutral situation; but, actually,
any shift $\hat{N} \rightarrow \hat{N}- \langle \hat{N} \rangle $
would produce a constant and a linear term in $\hat{N}$ that just
modifies $\m$.

Both  leads  are free electron gases with  chemical potentials
$\mu_{\gamma}$, $\gamma =l,r$;  hence
\begin{equation}
  H_{\rm leads}=\sum_{\g=l,r}\sum_{k,\s}
  (\varepsilon_{k}-\m_{\g})
  c^{\dagger}_{k,\g,\s}c_{k,\g,\s}
\end{equation}
with $\mu =\mu_{l}= \mu_{r}-eV$\cite{nota} where $V$ is the bias.

Finally the tunneling Hamiltonian is taken to be
\begin{eqnarray}
  H_{T} = \sum_{\eta, k,\s} && \left[
T^{(l)} (c^{\dagger}_{k,l,\s}f_{1, \eta \s} + {\rm h.c.})+ \right.
 \nonumber \\
&& \left.  T^{(r)} (c^{\dagger}_{ k,r,\s }f_{[\L/2], \eta \s} + {\rm h.c.}) \right]
\end{eqnarray}
where the $f^{\dagger}_{\a, \eta \s}$ are eigen-operators of the
noninteracting term of $H_{\a}$: $\sum_{i\s}t( p^{\dag}_{\a,0
\s}p_{\a, i\s}+ p_{\a, i\s}^{\dag}p_{\a,0\s}) = \sum_{\eta,\s}
\epsilon_{\eta} \, f^{\dag}_{\a, \eta \s} \, f_{\a,  \eta \s}$. We
observe that the tunnel junctions connect  two opposite clusters
to the leads; namely the $\a=1$ cluster is linked to the left
electrode and the  $\a=[\L/2]$ cluster to the right lead (here
$[x]$ means the integer part of $x$), see Fig.\ref{device}$a$. Note
also that $T^{(\g)}$ is independent of $\eta$, in other terms we
are using "white" wires, that is, leads that do not filter
electrons according to the square symmetry of electronic states in
each cluster. This is a simple way to ensure the 3D nature of the
leads, which is essential to allow Andreev reflection.

In the next Section we draw some relevant properties of $H_{\rm device}$
which mimic the behavior of $H_{BCS}$ despite the presence of strong electronic
correlations.

\section{Properties of the Central Island}
\label{sec2}

\begin{figure}
    \includegraphics[width=7cm]{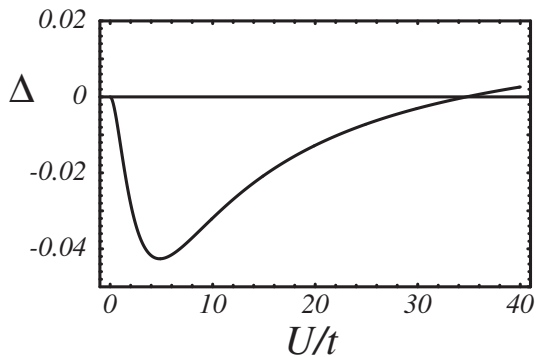}
    \caption{\label{delta}
     $\D$ (in eV) as a function of $U/t$.  The maximum
     binding occurs at $U \sim 5 t$ where
        $\D \approx -0.042\;t$.  For $U> 34 t$
        $\D$ becomes positive and pairing disappears.
   }
\end{figure}

In order to understand the physics of the device that we propose,
it is useful to foucus first  on properties of $H_{\a}$, referred
to a single 5-site cluster.   The Hamiltonian $H_{\a}$ is a
prototype example of electronic pairing from repulsion; this is
signaled  by the property $\D<0$, where $\D=\varepsilon(4)
+\varepsilon(2) -2 \varepsilon(3)$ and $\varepsilon(m)$ is the
ground state energy with $m$ electrons. There is pairing at $m=4$
 for $U/t<34$,  the minimum value is $\D \equiv -50$meV  at $t=1$eV
and $U \sim 5$eV and the binding energy is $| \D|$ (see Fig.\ref{delta}). The mechanism
has been investigated elsewhere\cite{PRB1997}\cite{EPJB1999}, and
need not concern us here; we just say that broadly speaking  it is
a lattice counterpart of the Kohn-Luttinger mechanism\cite{kohn}.

When  a negative $\D$ occurs, its competition with
$\frac{e^{2}}{C}$ determines the parity of the number $N_{gs}$ of
electrons in the ground state of $H_{\rm device}$ for small values
of the inter-cluster hopping $\t$.

Here we are interested in the behavior of the device at low
temperatures $k T \ll |\Delta|.$ Up to $o(\t)$ and $o(\t^{2})$
corrections, the ground state energy $E(N)$ of the central device
with fixed even or odd number of particles $N$ is
\begin{equation}
    E_{N} =
\left\{
\begin{array}{ll}
    \L  \varepsilon(2)
+(N-2\L)I_{P}+(\frac{N}{2}-\L) \D +  \\
\\
+\frac{(Ne)^{2}}{2C} -e(V_g-\mu) N \quad \quad \quad {\rm for \; \; even
\; \;}N  \\ \\ \\

    \L  \varepsilon(2)
+(N-2\L)I_{P} +(\frac{N}{2}-\L-\frac{1}{2}) \D +  \\ \\
+ \frac{(Ne)^{2}}{2C} -e(V_g-\mu) N \quad \quad \quad {\rm for \; \;
odd \; \;}N\; ,
\end{array}
\right.\;
\label{vgate}
\end{equation}
where $I_{P}=\varepsilon(3) -\varepsilon(2) $. Since one  bound
pair exists at $m=4$ electrons  in the 5-site cluster, the first
bound pair in the $\L$-cluster system appears at $N =2\L+2$
electrons. From Eq.(\ref{vgate}), it follows that in the range $ 2 \L
\leq N \leq 4 \L, $ $N_{gs}$ is always even if the pair binding
energy overcomes the charging energy. We call the situation when
$|\D|> \frac{e^{2}}{C}$  the {\em superconducting} regime.
Otherwise the system is {\em normal} and any $N$ is lowest in a
range of $V_{g}$.
\begin{figure}
    \includegraphics[width=7cm]{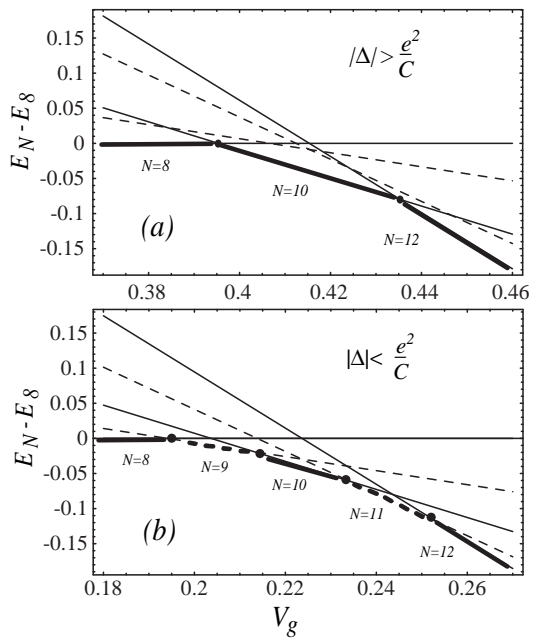}
    \caption{\label{even}
     $E_{N}-E_{8}$, where $E_{N}$ is the ground state energy  of $H_{\rm device}$, versus $V_{g}$, for
   several numbers $N$ of particles. $(a)$
    in the {\it superconducting} regime; $(b)$ in the in the {\it
    normal} regime.
    Solid lines are used for even $N$ and dotted lines for odd
    $N$. In the insets we plot the grand-canonical averages of $\hat{N}$
    in $H_{\rm device}$. We used $\L=4$, $t=1$eV,
    $\t=0$, $kT=0.001$eV, $C =50$e/V; with this choice $e^{2}/C=0.02$eV.
    In $(a)$ $U=5$eV ($\D=-0.043$eV ) and
    in $(b)$ $U=0.2$eV  ($\D=-0.0008$eV). $V_{g}$ is in V, $E_{N}$ is
    in eV.
   }
\end{figure}
In Fig.\ref{even} we plot  the ground
state energy of $H_{\rm device}$ as a
function of $V_{g}$ in both regimes, for $\t = 0$. It also
relevant to focus on the critical values of $V_{g}$ where
 ground states of different $N_{gs}$ cross. We define $(\D  V_{g})_{n}$ the spacing between the critical
values in the  normal  regime and   $(\D
 V_{g})_{sc} $ such a spacing in the  superconducting  regime;
it holds
\begin{equation}
\left\{
\begin{array}{ll}
   (\D  V_{g})_{n}  = \frac{e}{C} + \frac{\D}{e} \\
   \\
   (\D
V_{g})_{sc} = \frac{2e}{C}\;.
\end{array}
\right.\;
\label{parity}
\end{equation}
The charge fluctuations in the  superconducting regime as a
function of the gate voltage are about double spaced with respect
to the normal regime, that is the typical condition experimentally
realized in a S-SET\cite{devoret}.

\begin{figure}
    \includegraphics[width=7cm]{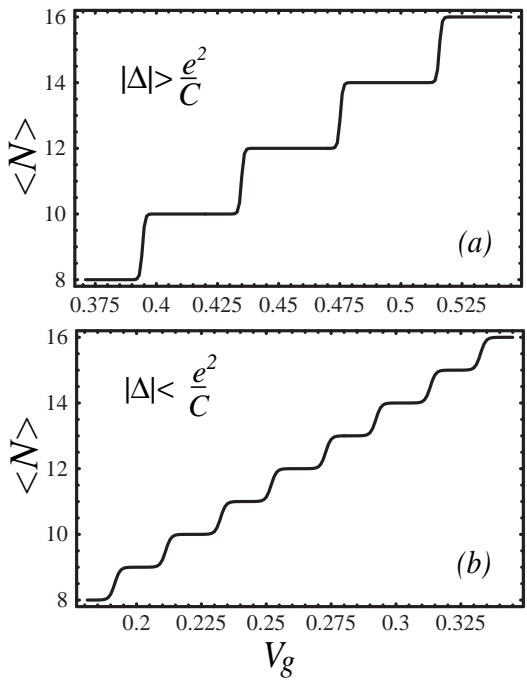}
    \caption{\label{staircase}
     $\langle N  \rangle$ versus $V_{g}$.$(a)$
    in the {\it superconducting} regime; $(b)$ in the in the {\it
    normal} regime.
    The parameters are the same as  in Fig.\ref{even}. $V_{g}$ is in V.
    }
\end{figure}

We can also visualize the previous results by plotting the gran-canonical
average of the number of particles in the isolated central device as
function of $V_{g}$. We use the standard definition
\begin{equation}
\langle N  \rangle = \frac{1}{Z} Tr [ \hat{N} e^{- (H_{\rm device}-\mu
\hat{N})/kT} ] \, ,
\label{averageN}
\end{equation}
where   $Z=Tr [e^{- (H_{\rm device}-\mu \hat{N})/kT}]$;  $Tr$ is
dominated by the  low energy states of  $H_{\rm device}$ with
$\t=0$. In Fig.\ref{staircase}$a$ one can observe the so called
Cooper staircase, characteristic of  the superconducting regime.

Below, for computational convenience, we assume
 $\t \ll |\D|$ and   deal with $H_{\t}$ perturbatively\cite{PRB2003}.
So, the critical values of $V_{g}$ where level crossing occur, are
spread into intervals of width $o(\t)$ in the normal regime and
$o(\t^{2})$ in the superconducting regime. Anyway the qualitative
behavior  is still close to Figs.\ref{even},\ref{staircase}.

\section{Calculation of the Linear Conductance}
\label{sec3}

Next, we consider the effects of small bias voltage $V$ applied
between leads, {\it i.e.} the linear conductance $G \equiv
\partial I / \partial V$ for $V \to 0$, versus the gate voltage
$V_{g}$. In the present  article, we follow the approach proposed
by Beenakker\cite{beenakker} and get  the formula for the
conductance  from a master equation. We  take $T^{(l)},T^{(r)} <<
kT << |\D |$ in order to provide that $(i)$ the parity of the
ground state is stable with respect to thermal effects, $(ii)$ the
elementary tunnel processes between the leads and the central
devices involve few particles at a time  and the broadening of the
levels of $H_{\rm device}$ due to the presence of the leads is
smaller than the thermal one. As discussed by
Beenakker\cite{beenakker}, these limitations characterize the
Coulomb blockade regime.

In the normal regime single-electron tunneling dominates. The
theory works very much like in Ref.\cite{beenakker} and we
calculate the first-order transmission rates
\begin{equation}
    \Gamma^{(\g)}_{q_{N-1},i_{N}} = \frac{2 \pi}{\hbar} T^{(\g)^{2}}
 \left| \sum_{\eta ,\s} \langle  q_{N-1}  |f_{\a,\eta \s} |i_{N}\rangle
 \right|^{2}
 \end{equation}
from the $i_{N-1}$-th state of the central device with $N$
particles (denoted $|i_{N}\rangle$) to the $q_{N-1}$-th state of
the central device with $N-1$ particles ( $|q_{N-1}\rangle$) via
tunneling to the left ($l$) or rigth ($r$) lead. Remember that $\a=1$
for $\g=l$ and $\a=[\L/2]$ for $\g=r$.
In the actual
calculations below, we obtain $|i_{N}\rangle$ in second-order
perturbation theory in $H_{\t}$; we mix the degenerate
ground-state multiplets of energy $E_{N}$ of $H_{\rm device}$,
which determine the low-energy properties of the system. By
first-order perturbation theory in $H_{T}$, one gets the familiar
formula:
\begin{eqnarray}
G^{(\rm I)}&=&\frac{ \r \,  e^{2}}{K T} \sum_{N=2\L+1}^{4\L} \; \;
\sum_{i_{N},q_{N-1}}
 \frac{\Gamma ^{(r)}_{q_{N-1},i_{N}}\Gamma^{(l)}_{q_{N-1},i_{N}}}
 {\Gamma^{(r)}_{q_{N-1},i_{N}}+\Gamma^{(l)}_{q_{N-1},i_{N}}}  \nonumber \\
 && \times P^{(0)}(i_{N}) [1-f(E_{i_{N}}-E_{q_{N-1}})]  \, .
\label{result1}
\end{eqnarray}
Here, $P^{(0)}(i_{N})$ is the Boltzmann equilibrium probability
for occupying the eigenstate $|i_{N}\rangle$ with energy
$E_{i_{N}}$;
 $f$ is the Fermi
distribution function, and $\r$ is the  density of states at the
Fermi level in both leads. Each term in Eq.(\ref{result1}) depends
on $V_g$ through the statistical factor $ P^{(0)}(i_{N})
[1-f(E_{i_{N}}-E_{q_{N-1}})] $ and     produces the well known
Coulomb blockade behavior\cite{beenakker}\cite{zhiming}. The
linear conductance is highly suppressed unless the gate voltage is
fine tuned at  $E_{i_{N}} \sim E_{q_{N-1}} - \mu$, where sharp
peaks of $G^{(\rm I)}$ occur. The second-order contribution in
$H_{T}$ depends on $o([T^{\g}]^{4}),$ rates which are negligible
with respect to the $o([T^{\g}]^{2})$   $\G$ coefficients;
therefore we can safely avoid working out the second-order current
in this regime.
\begin{figure}
 \includegraphics[width=7cm]{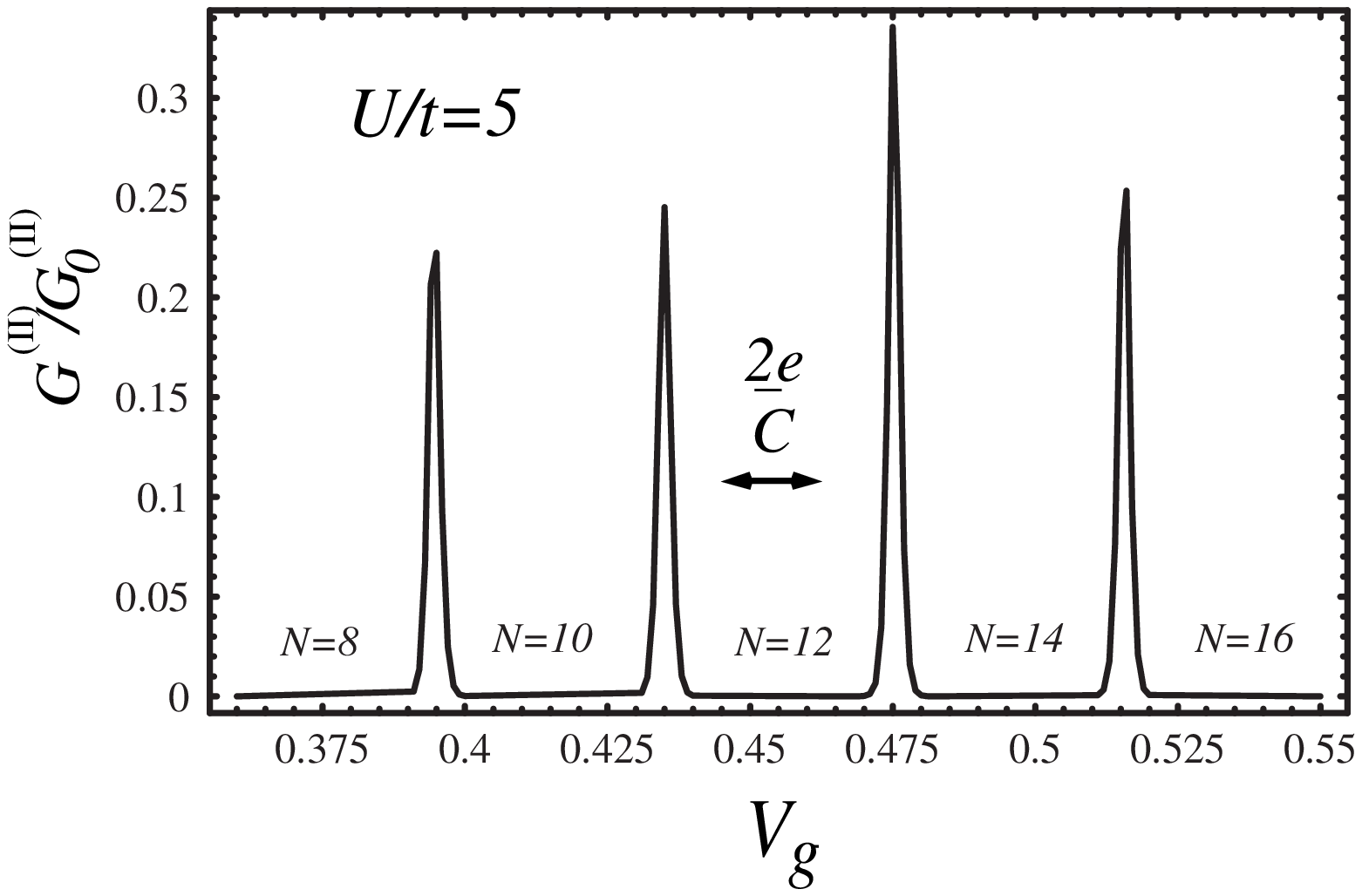}
    \caption{\label{suppeak} Linear conductance $ G^{\rm (II)}$  as a function of $V_{g}$
    in the pair-tunneling regime. We used $\L=4$, $t=1$eV,
    $U=5$eV, $\t=0.0005$, $kT=0.001$eV, $C=50$e/V, $\m$ =0. $V_{g}$ is in V.}
\end{figure}

When $|\D| > \frac{e^{2}}{C}$, only even $N$ have an important
weight in the appropriate range  of $V_{g}$ (see Fig.\ref{even}
$a$); therefore the resonance condition
$E_{i_{N}}=E_{j_{N-1}}-\mu$  never holds and the first-order
conductance $G^{(\rm I)}$ is highly suppressed for any value of
$V_{g}$. In this {\em pair tunneling regime,} accordingly, we must
go on calculating the  conductance up to second-order in
$H_{T}$\cite{averin}\cite{hekking}\cite{hanke}. Three-body,
four-body transitions and so on can be disregarded, however, as
 $T^{(l)}$ and $T^{(r)}$ are both small compared to the charging
energy. Since electrons can get paired in the device but not in
the leads, we may think of the second-order processes in terms of
Andreev reflections. First one of the two electrons tunnels from
one lead to the device (which is in the $|m_{N-2} \rangle$ state)
and forms a virtual excited state. Then the second one tunnels
into the device and form a bound pair ($|i_{N} \rangle $ state).

In principle in second-order, one should also take into account
the
 {\it cotunneling} processes\cite{averin2}, which  leave the
 population of the central island unchanged. Such processes
  provide a current away from the resonances.
 Anyway, as long as our device
 is in small bias and low temperature regime,
 the cotunneling current is found to be   negligible,
 as in the case of Ref.\cite{hekking}. Therefore the {\it sequential tunneling}
$m_{N-2} \to i_{N}$ is the major transport mechanism, and it is
possible only at the two-electron degeneracy points.
\begin{figure}
\includegraphics[width=7cm]{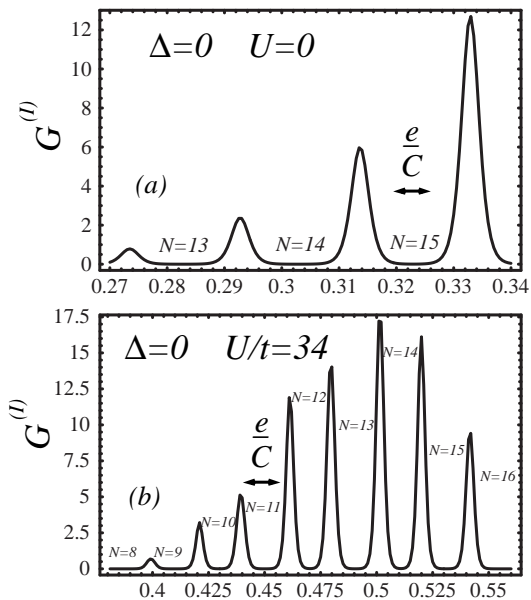}
    \caption{ \label{normpeak}Linear conductance $ G^{\rm (I)}$  as a function of $V_{g}$
    in the single-electron-tunneling regime: $(a)$ $U=0$,
    $(b)$ $U=34$eV. The other
parameters are the same as  in Fig.\ref{suppeak}. $V_{g}$ is in
V.}
\end{figure}
The full derivation  of the solution of  the detailed balance
equations\cite{beenakker} will be presented elsewhere. Setting
\begin{equation}\label{g0}
G_{0}^{\rm (II)}=\frac{128 \pi e^{2} \r^{2} }{\Delta^{2}\hbar}
 \frac{T^{(l)^{4}} T^{(r)^{4}}}{T^{(l)^{4}} +
  T^{(r)^{4}}}\;,
\end{equation}
one gets for the conductance
 \begin{eqnarray}\label{g2}
&&\frac{G^{\rm (II)}}{G_{0}^{\rm (II)}}=
 \sum_{N=2\L+2}^{4\L,\rm even}\; \sum_{i_{N},m_{N-2}}
   \Xi_{i_{N},m_{N-2}}e^{\frac{E_{i_{N}}-E_{m_{N-2}}- 2 \mu}{2kT}}\nonumber\\
   &&\times P^{(0)} (i_{N})
\frac{(E_{i_{N}}-E_{m_{N-2}}-2 \mu)/2kT}{{\rm
 sinh}[(E_{i_{N}}-E_{m_{N-2}}-2 \mu)/2kT]}
 \, .
\end{eqnarray}
The amplitude
\begin{eqnarray}\label{nova}
 \Xi_{i_{N},m_{N-2}}=
 \left| \sum_{l_{N^{-1}}}\sum_{\a, \b, \h ,\n
}
   \langle i_{N}
 |f^{\dagger}_{\a,\eta } | l_{N-1} \rangle  \langle l_{N-1} |
 f^{\dagger}_{\b,\nu } |
 m_{N-2} \rangle   \right|^{2}\nonumber
\end{eqnarray}
takes into account the second-order
 process governing the Andreev reflection.
Eq.(\ref{g2}) predicts   Coulomb blockade peaks for every $V_{g}$
such that  $E_{i_{N}}=E_{m_{N-2}}+2\mu$, while the conductance is
strongly suppressed elsewhere. Each peak has a correlation weight
due to the coefficient $\Xi_{i_{N},m_{N-2}} $, containing all the
microscopic information on the correlated ground states of the
central device.

For illustration, we numerically computed the conductance for a
central device with  $\L=4.$ In the superconducting regime,   $
G^{\rm (II)}$ as a function of the gate voltage is shown in
Fig.\ref{suppeak} for
 $t=1$eV, $U=5$eV, $\tau=0.0005$eV, $kT=0.001$eV, $C
=50$ e/. Note that $\tau \ll |\D|=0.043$eV, and $|\D| >
e^{2}/C=0.02 $eV.   $ G^{\rm (II)}$ shows neat  peaks, with
 spacing $(\D  V_{g})_{sc} =2e/C = 0.04$ V.

This superconductor-like behavior depends on the existence of
pairing. As a countercheck,   we calculate the linear conductance
$G^{(\rm I)}$ in the normal regime, when $\D=0$. We can obtain
this condition in two ways, namely in the  non-interacting case
when $U=0$ and in the very strongly correlated regime, when
$U\simeq 34$eV, the other parameters remaining the same as  in
Fig.\ref{suppeak}.
 $G^{(\rm I)}$  is plotted in Fig.\ref{normpeak}; since we are mainly interested in the
period of the resonances, we use constant $\G$'s and plot the
results in arbitrary units. Indeed,  for $\D=0$ the period of the
resonant peaks is $e/C \simeq 0.02$V, {\it i.e.} a half of the
period in the superconducting case.

\section{Summary and Conclusions}
\label{sec4}

We have shown that the  Hamiltonian ${\cal H}$ models a S-SET in
the linear regime. We pointed out that the repulsion-induced
pairing occurring in $H_{\rm device}$ fixes a characteristic
energy $|\D|$ which competes with the electrostatic charging
energy $e^{2}/C$. As in any  S-SET,  there is   a {\it normal
regime}, where $|\D|<e^{2}/C$ and a {\it superconducting regime},
where $|\D|>e^{2}/C$. In the first case the parity of the electron
number  in the ground state oscillates between even and odd values
and the transport properties are governed by single-electron
tunneling. Conversely in the superconducting regime the parity is
always even and the major transport mechanism  is sequential
tunneling of pairs. The explicit calculations have been performed
for a ring of four 5-site clusters, but a general expression for
the linear conductance is also derived.

Our results suggest a systematic way to produce a well controlled
periodic two-electron pattern, even without any  conventional
superconductivity;  an array of quantum dots similar to the one in
Ref. \cite{ashoori} could be designed to this purpose.

Finally we underline that the model we propose is very flexible
with respect $(i)$ to the size and the shape of the Hubbard
clusters, $(ii)$ to the topology of the cluster array forming the
central device. Indeed a wide variety of Hubbard clusters show the
$\D<0$ property at proper fillings, which is actually the key
feature at the basis of our device; we could construct many
alternative devices, based on graphs with different topologies,
also in view of possible single-electronics applications to more
complex circuits than a transistor.




\begin{references}
\bibitem{definizione} In this case the device is usually  denoted by
NSN-SET, in order to distinguish it from SSS-SET, where the
superconducting island is connected to superconducting leads;
anyway in this article we use S-SET for short.
\bibitem{nota} With reference to Eq.(\ref{model}), the total
number of particles is $\hat{N}_{tot}=\hat{N}+\sum_{\g,k,\s}
c^{\dagger}_{k,\g,\s}c_{k,\g,\s}$.
\end{references}
\end{document}